\documentclass[11pt]{article}
\usepackage{helvet}
\usepackage{url}
\usepackage{a4,graphicx}
\usepackage{natbib}
\usepackage[latin1]{inputenc}
\usepackage{amssymb}
\usepackage{amsmath}
\usepackage{paralist}
\usepackage{bibspacing}
\usepackage{chapterbib}

\usepackage{mathrsfs}
\usepackage{wasysym}
\usepackage{array}
\usepackage{makeidx}
\usepackage{floatflt}
\usepackage{dcolumn} 
\usepackage{multicol}
\usepackage{ifthen}
\usepackage{eurosym}
\usepackage{sidecap}
\usepackage{hyperref}
\renewcommand{\refname}{3~~~Project- and subject-related list of publications} 

\usepackage{xspace}
\usepackage{color}
\usepackage{amsmath}

\newcounter{Rco}
\newcommand{\Ionst}[1]{\setcounter{Rco}{#1}\Roman{Rco}}
\newcommand{\Ion}[2]{\mbox{#1\,{\scriptsize\Ionst{#2}}}}

\newcommand{\Teff}{\mbox{$T_\mathrm{eff}$}\xspace}

\begin{document}

\begin{center}

{\Large \bf Building a Roadmap for Hubble science into the 2030s - White Paper}
\vspace{0.5cm}

{\Huge \bf Studying hot evolved stars with ultraviolet spectroscopy}\\
\vspace{1.0cm}

{\Large Stephan Geier$^1$, Nicole Reindl$^2$, Matti Dorsch$^1$, Vikrant Jadhav$^3$, Helge Todt$^1$, Klaus Werner$^4$, Ulrich Heber$^5$, Marcelo M. Miller Bertolami$^{6,7}$, Tiara Battich$^{6}$, Semih Filiz$^2$}

\end{center}
\vspace{0.5cm}
{\small $^1$Institut f\"ur Physik und Astronomie, Universit\"at Potsdam, Haus 28, Karl-Liebknecht-Str. 24/25, D-14476 Potsdam-Golm, Germany}\\
{\small $^2$Zentrum f\"ur Astronomie der Universit\"at Heidelberg, Landessternwarte, K\"onigstuhl 12, D-69117 Heidelberg, Germany}\\
{\small $^3$Astronomical Institute, Faculty of Mathematics and Physics, Charles University, V Hole\v{s}ovi\v{c}k\'ach 2, Praha, 18000, Czech Republic\\
{\small $^4$Institut f\"ur Astronomie und Astrophysik, Universit\"at T\"ubingen, Sand 1, D-72076 T\"ubingen, Germany\\}
{\small $^5$Dr. Remeis-Sternwarte \& ECAP, University of Erlangen-N\"urnberg, D-96049 Bamberg, Germany\\}
{\small $^{6}$Instituto de Astrof\'{i}sica de La Plata, Consejo Nacional de Investigaciones Cient\'{i}ficas y T\'{e}cnicas, Avenida Centenario S/N, La Plata B1900FWA, Argentina\\}
{\small $^{7}$Facultad de Ciencias Astron\'{o}micas y Geof\'{i}sicas, Universidad Nacional de La Plata, Avenida Centenario S/N, La Plata B1900FWA, Argentina}

\paragraph{Abstract:} Hot evolved stars are key objects to reconstruct the various evolutionary pathways of Sun-like stars, to probe binary interactions and the physics of supernovae. They serve as powerful observational constraints to test diffusion, mixing, and mass loss in hot stellar atmospheres. Furthermore, hot stars serve as laboratories to test and derive atomic data for highly ionised trans-iron group elements and to investigate different nucleosynthesis models. Hot evolved stars emit most of their flux in the ultraviolet (UV) and a lot of progress has been made in characterizing their UV-spectra both on the observational and on the modelling side. The unique capabilities of HST to obtain high- and medium-resolution UV-spectra played a crucial role and are needed to further advance this field also in preparation for HWO.


\paragraph{The importance of hot evolved stars for astronomy and physics:}
The region of the hot subluminous stars and hottest white dwarfs (WDs) located in luminosity between the main sequence and the WDs in the diagnostic Hertzsprung-Russell diagram remains as one of the least understood despite the fact that no less than $97\%$ of all stars are evolving through it at some point during their lives. This is due to the short evolutionary times or peculiar evolutionary channels of the stars populating this region. New types of stars with unclear origins and evolutionary states are still discovered. Most stars are entering the hot subluminous regions after passing the asymptotic giant branch (AGB), but they evolve so fast that only relatively few objects are actually observed in the post-AGB phase.

A significant fraction of hot evolved stars have atmospheres almost entirely devoid of H and sometimes even He is absent. These H-deficient stars pose a major challenge to canonical stellar evolution theory as they cannot be reproduced by standard single-star evolutionary models making them unique laboratories for probing peculiar and otherwise inaccessible stages of stellar evolution. Importantly, H-deficient objects constitute a significant fraction (up to $\sim20$\%) of all WDs, the most common endpoints of stellar evolution and key tracers of Galactic structure and evolution \citep{Isern+2022}.

H-deficient atmospheres may form either through violent internal processes, such as late He-shell flashes, or through binary interactions, particularly double-degenerate mergers. Consequently, hot H-deficient stars provide powerful constraints on merger channels and on the formation of ultra-massive and highly magnetic WDs \citep{Reindl+2014b, WernerRauch2015, Werner+2022b, miller-bertolami2022, Mackensen+2025}. Moreover, proton ingestion events during late He-flashes are the preferred sites of intermediate neutron-capture processes and, therefore, play an important role in our understanding of the formation of trans-iron elements \citep{2011ApJ...727...89H,2025NatRP...7..696W, battich25}. 

Binary interactions are also the likely origin of other types of hot subluminous stars \citep[sdO/Bs,][]{heber16}. Because their envelopes are stripped by mass-transfer to a close companion before or at the tip of the first giant phase, they either end up in a core helium-burning phase (extreme horizontal branch, EHB) or they ignited no He-fusion at all and evolve to become He-WDs. 
In addition, merger scenarios involving He- or CO-WDs in tight binaries have been invoked to explain the formation of diverse types of single hot evolved stars \citep{zhang12, Reindl+2014b, Werner+2022b, WernerRauch2015, miller-bertolami2022}. 

Close binary sdO/Bs are among the best candidates for the progenitors of type Ia supernovae (SN\,Ia) and sources for gravitational wave radiation detectable with space-based instruments \citep{geier13,pelisoli21}. Additionally, fast moving hot subluminous stars and WDs may be surviving remnants of supernova explosions, ejected at high velocities from their birth sites \citep{Geier+2015b, Werner+2024a}. Studying and characterizing these runaway objects provides an alternative window into studying supernova mechanisms and the disruption of binary systems. Their occurrence rates may help quantify the role of different double-degenerate pathways in producing SN\,Ia, while their atmospheric compositions can offer valuable clues about nucleosynthesis processes.

The study of hot evolved stars is also crucial to understand the formation of the diverse shapes of planetary nebulae, which is inevitably linked to the evolution of the hot central stars \citep{Jones+Boffin2017}. Some centrals stars have even been observed to evolve in "real-time" (e.g. \citealt{Reindl2014, Reindl+2017}), meaning that changes of the surface properties of a given star occurs on a time scale shorter than a human life-time, and can thus be witnessed by careful observations performed at different times. This provides a unique way to gain a direct knowledge of stellar evolution.

Hot evolved stars and WDs are of importance for other fields of astrophysics and physics general: Hot WDs are considered the most reliable and internally consistent flux calibrators \citep{Bohlin+2020}. Hot WDs and sdO/Bs serve as valuable laboratories to derive atomic data for highly ionized species of trans-iron elements and their chemically peculiar atmospheres are also well suited to study mixing, mass loss, and diffusion processes \citep{Rauch+2015a,dorsch26}. The diverse types of pulsators among the hot evolved stars can be studied by asteroseismic modelling \citep[e.g.][]{charpinet19}. Newly formed WDs allow us to study the age structure and star formation history of the Galactic halo \citep{Kalirai2012}, while sdO/Bs are regarded as important contributors to the UV-excess seen in early-type galaxies \citep{lisker2008} and are crucial to understand the HB morphology and evolution of globular clusters \citep{piotto15}. Stripped stars of intermediate mass are regarded as important to explain the reionization of the Universe \citep{goetberg17}. The shape of the WD luminosity function is sensitive to the properties of proposed weakly interacting particles \citep{MillerBertolami2014b}. Hot WDs also potentially allow us to directly observe variations in fundamental constants -- like the fine structure constant -- at locations of high gravitational potential. Such a variation manifests itself as shifts in the observed wavelengths of spectra lines compared to laboratory wavelengths \cite{Berengut+2013}.  



\paragraph{The importance of UV observations to study hot evolved stars:}

Hot evolved stars radiate the bulk of their energy in the ultraviolet (UV). Already the first dedicated surveys to search for these objects used near-ultraviolet bands \citep{humason47}. Diverse UV surveys from space played an important role in extending the sample early on  \citep{vanduinen75,wesselius76,laget78,carnochan83,viton88,brosch95}. Although state-of-the-art catalogues of hot subluminous stars are now compiled based on Gaia data, a combination with UV photometry is still advantageous, especially when searching for UV-excesses in composite systems \citep{dawson24}. 

The HST UV legacy survey had an unprecedented impact on studying globular cluster stars because of HST's high spatial resolution. HST UV photometry allowed finding rare hot stars of diverse types at ''inappropriate'' positions in the HRD, including sdO/Bs and other UV-bright post-AGB stars \citep{2017AJ....154..126D,2024ApJ...961...24D}. Given their well understood population properties, the study of post-AGB stars in globular clusters allows for unique insights into possible evolutionary channels and the past nucleosynthesis of these stars \citep{2019AJ....157..147D,2021AJ....162..126D, 2025ApJ...980....5D}.   

Pointed observations with the UVIT telescope on board of the ASTROSAT observatory \citep{subramaniam16} and the UVOT telescope onboard Swift observatory \citep{gehrels04} are used to detect UV excesses  caused by optically invisible hot subluminous stars in cluster binaries. In a similar way, UV photometry with HST recently allowed us to detect a hidden population of intermediate-mass sdOs in the Magellanic Clouds \citep{drout23}.

The era of UV spectroscopy of sdO/Bs started with Copernicus and IUE, which allowed for the first time pointed high- and low-resolution spectra to be obtained for a substantial number of objects. The first quantitative spectral analyses in the UV were performed using IUE data \citep{heap78,heber84} and led to several key discoveries. The first metal abundance patterns determined from UV-lines \citep{baschek80,lamontagne85} confirmed sdO/Bs to be chemically peculiar with diffusion processes to be proposed to explain this phenomenon. Mass loss by stellar winds in luminous sdOs was first detected in the UV \citep{hamann81} as well as the first binary consisting of an sdO and a Be type star \citep{thaller95}. It became clear that constraints from UV spectra are needed to improve model atmospheres and determine reliable temperatures of the hottest sdO/Bs \citep{werner96}. 

FUSE spectroscopy extended the available wavelength range downwards and allowed to measure the abundances of more and heavier elements \citep{ohl00}. The improved sensitivity of HST, however, allowed for more detailed studies, also of  fainter stars, already starting with the GHRS spectrograph \citep{lanz97,gies98}. STIS spectra were used to show that the surface abundances of iron-group elements do not influence the pulsation properties of sdBs \citep{otoole06} and to provide the full abundance patterns of super-metal-rich He-sdOBs and hot WDs and to improve atomic data for iron and trans iron-group elements \citep{landsdorfer24, Rauch+2015a, dorsch26}. COS spectra were used to perform the first detailed studies of sdO/Bs in globular clusters \citep{chayer14,latour17}. For a more detailed review on UV observations of hot evolved stars see \citet{fu26}.

\paragraph{The need of UV spectroscopy of hot evolved stars:}

UV spectroscopy is indispensable for understanding the physical properties and evolutionary state of hot evolved stars. At effective temperatures often exceeding 50\,000 K, the strongest diagnostic lines of highly ionized species, such as \Ion{C}{3}-{\scriptsize IV}, \Ion{N}{3}-{\scriptsize V}, \Ion{O}{3}-{\scriptsize VI}, \Ion{Ne}{7}, and Fe-group elements fall primarily in the UV, where they provide crucial constraints on atmospheric composition, temperature, surface gravity, mass-loss rates, and stellar winds. Optical spectra alone are frequently insufficient because many key ions are weak or absent at visible wavelengths in these extreme atmospheres. High-quality UV spectra are therefore essential to derive metal abundances for these stars. 

Furthermore, effective temperatures and surface gravities of hot WDs and other hot subluminous stars derived from the Stark broadened optical Balmer and/or He II lines can suffer from huge systematic uncertainties. More than half of the H-rich (pre-)WDs with \Teff$\geq50$\,kK are affected by the so-called Balmer line problem \citep{Bedard2020}. It describes the failure to achieve a consistent fit to all Balmer lines simultaneously, meaning that for a particular object different \Teff follow from fits to different Balmer line series members. Analogous to the Balmer line problem, approximately one in four of the He-dominated WDs display the \Ion{He}{2} line problem \citep{Bedard2020}, where the observed \Ion{He}{2} lines are too broad and too deep compared to photospheric models. It has to be stressed that the standard method of constraining \Teff via fitting of spectral energy distributions (SEDs) is not applicable for hot stars, because for \Teff $\gtrsim 40$\,kK the shape of the SED is no longer sensitive to changes in \Teff and much more dependent on metal opacities or interstellar reddening. The only reliable way to constrain the \Teff of the hottest evolved stars is by exploiting metal ionization balances in UV spectra, which provide a much more precise \Teff indicator than the \Ion{H}{1} / \Ion{He}{2} line analysis. Several studies that employ the UV metal-ionization balances technique showed that systematic \Teff errors of $10-30$\% -- or even more -- are the rule rather than the exception, \citep[e.g.][]{Reindl+2014b, Werner+2019}, if UV spectra are not available.  

In many cases the optical spectra of hot (pre-)WDs are also contaminated by either a (compact) planetary nebula or a luminous companion which makes an optical spectral analysis difficult or even impossible. Finally, wind parameters, such as the mass-loss rates and terminal wind velocities, can only be determined in the UV \citep{2011MNRAS.417.2440H}.




\paragraph{Scientific objectives:}

A lot of progress has been made in characterizing UV-spectra of hot evolved stars both on the observational and on the modelling side. The unique capabilities of HST to obtain high-resolution UV-spectra are crucial to further advance this field and tackle in particular the following objectives:

\begin{itemize}


\item Detecting all of the hot evolved stars in different Galactic and extragalactic populations such as open and globular clusters, the bulge, the halo, or satellite galaxies is necessary to put constraints on the frequency of diverse types of interactions. It is predicted by theory and has been shown by observations that the binary properties of sdO/Bs differ between these populations \citep{geier24}. Studying these differences will allow us to understand the underlying interactions better. For this we need not only deeper photometric UV surveys with higher spatial resolution such as UVEX \citep{2021arXiv211115608K} or Ultrasat \citep{2024ApJ...964...74S}. HST spectroscopy is the only way to characterize hot evolved stars in binaries with early-type main-sequence companions that dominate at optical wavelengths, leaving the evolved stars detectable only through weak UV excesses. The high spatial resolution of HST is also essential to obtain the spectra of hot subluminous candidates outside the Milky Way, e.g., the Magellanic clouds \citep{2026ApJ...999...73L}.

\item The UV is the only wavelength range where abundances of heavy and even trans-iron elements are detectable in hot evolved stars, and the best to determine their metal abundances in general. These abundance patterns give important clues for potential formation channels, where episodes of stripping or mixing are involved. Very recently, it has been shown that the abundance patterns of some peculiar He-sdOBs are consistent with predictions from i-process nucleosynthesis models \citep{battich25,dorsch26}. Similarly, most H-deficient post-AGB stars (PG1159, and [WR] spectral types) are understood to be formed during late-helium flashes were i-processes are expected to occur \citep{2011ApJ...727...89H, 2024Galax..12...83M}. If confirmed, this will not only help to constrain the evolutionary history of these stars but might have much wider implications for the origin of the heavy elements in the Universe. Also circumstellar material, left over from envelope ejections \citep{li22} or mergers, can best be studied in the UV.

\item The atmospheres of hot evolved stars show many peculiarities, which make them ideal laboratories to study hot, radiative atmospheres in general. Their compositions can be so extreme that the structure of the atmosphere is changed significantly, radiative levitation and gravitational settling play a role \citep{Filiz+2024}, vertical stratification and Zeeman splittings have been observed, some sdOs show weak and potentially metallic winds, and many of the spectral features in the UV still remain unidentified due to the lack of atomic data. Hot WDs have also been employed over a decade to derive atomic data of trans-iron elements, offering valuable spectroscopic tools to investigate heavy elements in hot stars in general (e.g. \citealt{Rauch+2015a, Rauch+2017b}). Hot evolved stars are therefore ideal testbeds to improve models for hot stellar atmospheres and have been used to that end ever since. 

\end{itemize}

\paragraph{Prospects for HST in the 2030s:}

Access to HST UV spectrographs covering $900-2000$\,\AA\ with sufficiently high spectral resolutions ($\Delta \lambda \approx 0.1$\,\AA) to resolve the spectral features is mandatory to make further progress along those lines. The TIME-TAG mode is needed to resolve variations due to pulsations and binary-induced variability frequently observed in hot evolved stars.

In addition, archival UV spectra of those stars have a particularly high legacy value. The archival collections of STIS and COS are still in use to test improved models. The observations of stars evolving in ``real-time" are of profound legacy importance, with each observation contributing to a long-term archival record whose scientific value grows as analytical methods improve, enabling future retrospective studies of evolutionary phases that may themselves last longer than most scientists' careers. It is therefore crucial to keep these collections publicly available and well curated. 

It has also to be pointed out that it is mostly the lack of atomic data for many species that limits the diagnostic value of the spectroscopic data we have at hand. Theoretical and experimental research in this direction needs to be maintained and extended to make further progress. To establish and maintain connections between research groups in atomic and astrophysics \citep[see][for most recent results of such a collaboration]{dorsch26}, we need to be able to obtain high quality data in the future.

While detailed studies of hot evolved stars are currently only possible for relatively bright and close-by stars, HWO will allow us to extend these studies to other Galactic and extragalactic populations. New observing modes such a polarimetry can be used to study magnetic fields observed in some subtypes. The still growing diversity of hot evolved stars, where new discoveries are constantly made, calls for a characterization of all these diverse types with HST to prepare more focused observation with HWO. Also because deep multi-band surveys such as LSST will detect new faint blue candidates, which can only be followed-up spectroscopically with the next generaton of large UV telescopes. The recently completed STIS/COS treasury program 17697 observed 37 sdO/Bs and the STIS/COS large program 17112 recorded 111 of the hottest WDs. Both programs cover diverse types and mark an important step in this direction, but need to expanded in the future. 
 
\footnotesize
\bibliographystyle{aa}
\bibliography{hst2026_geier.bib}

\end{document}